\documentclass[useAMS,usenatbib]{mn2e}
\usepackage{graphicx}
\usepackage{times}

\begin{document}

\renewcommand{\topfraction}{1.0}
\renewcommand{\bottomfraction}{1.0}
\renewcommand{\textfraction}{0.0}

\newcommand{\Avg}[1]{{\langle}{#1}{\rangle}}
\newcommand{\Cn}[0]{\mbox{$C_n^2$}~}
\newcommand{\bfit}[1]{\mbox{\boldmath $#1$}}
\newcommand{\bfsf}[1]{\mbox{\boldmath \sf #1}}

\title[Combined MASS-DIMM instrument]{Combined MASS-DIMM instrument for atmospheric turbulence studies}

\author[Kornilov et al.]{V.~Kornilov$^{1}$\thanks{E-mail:
victor@sai.msu.ru}, A.~Tokovinin$^2$, N.~Shatsky$^{1}$, O.~Voziakova$^{1}$, S.~Potanin$^{1}$, B.~Safonov$^{1}$ \\
$^1$Sternberg Astronomical Institute, Universitetsky prosp. 13, 119992 Moscow, Russia \\
$^2$Cerro Tololo Inter-American Observatory, Casilla 603, La Serena, Chile
}

\date{-}

\pagerange{\pageref{firstpage}--\pageref{lastpage}} \pubyear{2007}

\maketitle

\label{firstpage}

\begin{abstract}

 Several  site-testing   programs  and  observatories   currently  use
combined  MASS-DIMM instruments for  monitoring parameters  of optical
turbulence.  The  instrument is described here.  After  a short recall
of the measured quantities  and operational principles, the optics and
electronics of MASS-DIMM, interfacing  to telescopes and detectors,
and  operation are covered  in some  detail.  Particular  attention is
given  to   the  correct  measurement  and   control  of  instrumental
parameters  to ensure  valid  and well-calibrated  data,  to the  data
quality and filtering.  Examples of MASS-DIMM data are given, followed
by the list of present and future applications.
\end{abstract}


\section{Introduction}

Light  propagation  in terrestrial  atmosphere  is  one  of the  major
factors limiting the performance of ground-based astronomy at optical,
infrared,  and  radio  wavelengths.   Most observatories  monitor  the
optical quality of the  atmosphere, {\em seeing}, while their location
has  been  usually selected  to  provide  good  seeing conditions.   A
classical  method of  measuring seeing  is based  on  the differential
motion of stellar images  formed by two small apertures.  Differential
Image  Motion Monitor  (DIMM) has  become  a {\em  de facto}  standard
instrument \citep{DIMM,PASP02}.

Nowadays,  knowledge  of  seeing  is  not  sufficient  because  modern
observing  methods such  as  adaptive optics  (AO) and  interferometry
depend  on additional  atmospheric parameters  -- {\em  time constant}
$\tau_0$ and {\em isoplanatic angle} $\theta_0$ \citep{Hardy}.  Vertical
turbulence profile (TP)  must be known to evaluate  the performance of
advanced  AO systems.  Even  for classical  astronomy, just  measuring
seeing is not enough because ultimate limits of precise photometry
and astrometry depend on the TP and wind-speed profile \citep{Kenyon}.

The  need  for a  better  characterisation  of atmospheric  turbulence
stimulated   development  of   new  methods   and   instruments,  e.g.
Generalised   Seeing   Monitor   \citep{GSM}  or   Single-Star   Scidar
\citep{Habib}.  A simple and practical way to measure low-resolution TP
and time constant by analysis of scintillation has been implemented in
a  Multi-Aperture Scintillation  Sensor (MASS)  \citep{MASS,Rest}. When
combined with DIMM in a  single instrument, it becomes a powerful tool
for advanced  site monitoring  and testing.  Several  such instruments
have been built and have already produced useful results.

We  feel it  timely to  give a  concise description  of  the MASS-DIMM
instrument and  its operation.  Particular emphasis is  placed  on
correct  calibration  and data  quality  control  to ensure  un-biased
estimates  of atmospheric  parameters.  The  biases of  DIMM  and MASS
methods are  studied in-depth by  analytical theory and  simulation in
the  accompanying paper \citep[][hereafter  TK07]{TK07}.  We  begin by
recalling  the  principles  of  MASS  and  DIMM.  Then  the  MASS-DIMM
instrument is  described in Sect.~\ref{sec:MASSDIMM}  and correct ways
to   set  and   control   instrument  parameters   are  discussed   in
Sect.~\ref{sec:use}.   Examples of the  applications of  the MASS-DIMM
and some recent results are given in Sect.~\ref{sec:examples}.

\begin{figure*}
\centerline{\includegraphics[width=14cm]{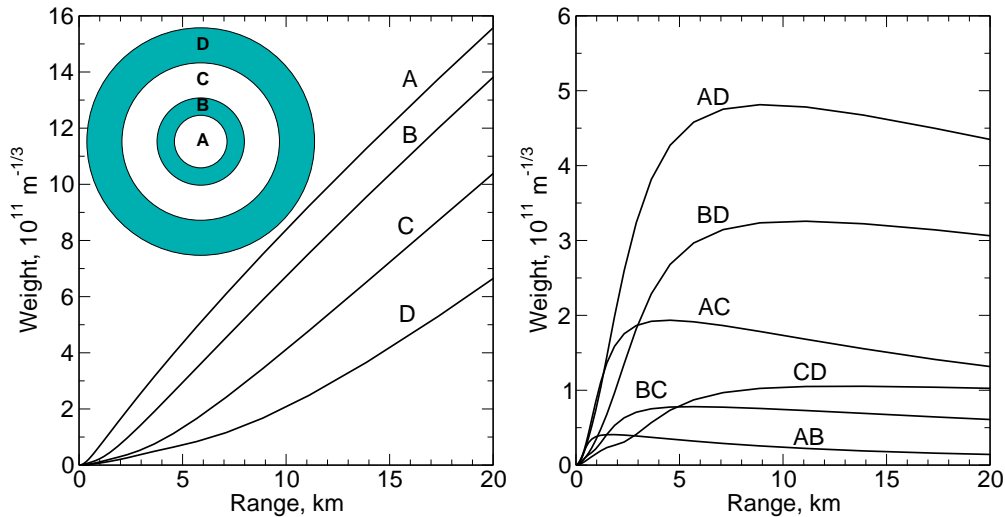} }
\caption{Weighting  functions of MASS 
  $W_k(z)$ corresponding to the  4 normal scintillation indices (left)
  and  6  differential   indices  of  pairwise  aperture  combinations
  (right).  The geometry of 4 annular apertures A,B,C,D with diameters
  1.9, 3.2, 5.6, 8~cm is shown in the left panel.
\label{fig:massweight} }
\end{figure*}

\section{Recall of the principles}
\label{sec:principles}

\subsection{Atmospheric parameters}

The  theory of light  propagation through  optical turbulence  is well
developed   \citep{Tatarskii,Roddier81}.    The   local  intensity   of
turbulence  is characterised  by the  {\em refractive  index structure
constant} \Cn.  The turbulence profile  (TP) $\Cn (h)$ is defined as a
function of  altitude above  observatory $ h  = z \cos  \gamma$, where
$\gamma$  is the  zenith  angle, $z$  is  the propagation  distance
(range).

The  atmospheric  image  quality,  {\em  seeing}, is  related  to  the
integral of  \Cn over  propagation path, $J$: 
\begin{equation}
J  = \int_{\rm path}  \Cn (z) {\rm d} z .  
\label{eq:J}
\end{equation}
Seeing is usually quantified by the {\em Fried parameter}
$r_0$, $r_0^{-5/3} = 0.423 (2  \pi/\lambda)^2 J$, or by the theoretical
FWHM of a point-source image $\varepsilon_0 = 0.98 \lambda/r_0$.  Both
$r_0$   and   $\varepsilon_0$  depend   on   the  imaging   wavelength
$\lambda$. To  avoid ambiguity,  it is recommended  to use  $\lambda =
500$\,nm  as  standard  and   to  reduce  all  parameters  to  zenith,
considering  that $r_0  \propto  ( \cos  \gamma )^{3/5}$.   Additional
parameters  of   interest  to  AO  and  interferometry   are  the  {\em
atmospheric time  constant} $\tau_0 = 0.31  r_0/\overline{V}$ and {\em
isoplanatic  angle}   $\theta_0  =  0.31   r_0/  \overline{H}$,  where
$\overline{V}$  and  $\overline{H}$  are,  respectively,  \Cn-weighted
average wind speed and altitude above site \citep{Roddier81}.
These definitions imply that optical turbulence is a stationary random process with a Kolmogorov spectrum.

\subsection{MASS: Multi-Aperture Scintillation Sensor }

The MASS instrument is based on the analysis of stellar scintillation.
The {\em scintillation index} $s^2_A$ is defined as
\begin{equation}
s^2_A = \Avg{ \Delta I_A^2} / \Avg{I_A}^2  ,
\label{eq:sig_I}
\end{equation}
where $I_A$ is the instantaneous light intensity received through some
aperture A, $\Delta I_A$ is its fluctuation.  The spatial scale of the
scintillation  ``speckle''  is of  the  order  of  the Fresnel  radius
$\sqrt{\lambda  z}$,  i.e.   $\sim  10$\,cm for  a  10-km  propagation
\citep{Roddier81}.   An aperture  of diameter  $D$ acts  as  a spatial
filter, admitting only fluctuations  with the spatial scales larger than
$D$.  With aperture diameter comparable  to the Fresnel radius, we can
distinguish  the altitude  where the  scintillation was  produced.  An
even better  method is to measure the  {\em differential scintillation
index} $s^2_{AB}$ between two apertures A and B,
\begin{equation}
s^2_{AB} = \left\langle \left(
\frac{\Delta I_A}{\Avg{I_A}} - 
\frac{\Delta I_B}{ \Avg{I_B}}   \right)^2\right\rangle . 
\label{eq:sig_d}
\end{equation}

In  the framework of  the small-perturbation  theory, both  normal and
differential  scintillation indices $s^2_k$  depend on  the turbulence
profile $\Cn (z)$ linearly as
\begin{equation}
s^2_k = \int W_k(z)\; \Cn (z) {\rm d} z ,
\label{eq:WF}
\end{equation}
where  the  {\em  weighting  function}  (WF)  $W_k(z)$  describes  the
altitude  response of  a given  aperture or  aperture  combination $k$
\citep{AO02,Poly}. If it is  constant, it means that the scintillation
index gives a direct measure of the turbulence integral, hence seeing.
The differential  index in two  concentric annular apertures  has this
attractive  property for  the propagation  distance  $z> D^2/\lambda$,
where $D$ is the  average diameter of the apertures \citep{AO02,Poly},
and thus provides  a more-or-less direct measure of  the seeing in the
free  atmosphere (Fig.~\ref{fig:massweight}).  With  several apertures
that match Fresnel  radii for a range of altitudes,  it is possible to
get   a   crude   estimate   of   the   TP.    Departures   from   the
weak-scintillation  theory underlying  the  formula (\ref{eq:WF})  are
studied in TK07 and accounted for in the TP restoration.

The  MASS has  4 apertures  and measures  10 scintillation  indices (4
normal and 6 differential).  This data vector {\bfit s} is fitted with a
model of 6  thin turbulent layers at altitudes $h_i$ of  0.5, 1, 2, 4,
8, and 16\,km with turbulence integrals $J_i$ as free parameters, 
\begin{equation}
{\bfit s} = {\bfsf W} \; {\bfit J},
\end{equation}
where {\bfsf W} is the $10 \times 6$ matrix of weights and ${\bfit J}$
is  a  6-element  vector   of  the  TP  with  non-negative  elements
\citep{Rest}.   Note that  the  zenith angle  $\gamma$  is taken  into
account because  the rows of the  matrix {\bfsf W}  are calculated for
$z_i =  h_i \sec \gamma$.   In reality, the turbulence  is distributed
continuously  in altitude  with a  profile $\Cn  (h)$.   The integrals
delivered  by MASS  are approximately  equal to  $J_i =  \int  \Cn (h)
R_i(h) {\rm d}h$, where the {\em response functions} $R_i(h)$ resemble
triangles in $\log h$  coordinate centred on $h_i$ \citep{Rest}.  The
sum of all $R_i(h)$ is close to one for $h > 0.5$\,km.

The restoration  algorithm described by \citet{Rest}  will be slightly
modified  when  two  additional  datums from  DIMM  (longitudinal  and
transverse  variance) are  available. These  numbers will  be included
into  the data  vector {\bfit  s}, and  the matrix  {\bfsf W}  will be
modifed accordingly by adding two rows with WFs for the DIMM response,
nearly constant  with $h$  (cf.  TK07). The  TP model will  contain an
additional layer  at $h=0$ and can  be fitted to the  combined data in
the  same way  as  the 6-layer  model.  This modification  is not  yet
implemented in the current software.

The scintillation is mostly  produced by high layers and, therefore,
the accuracy of MASS TPs is best for the highest layers and quite poor
for the lowest  ones. Random errors of $J_i$  are typically about 10\%
of  $\sum_i J_i$.   MASS has  been compared  to the  SCIDAR turbulence
profiler \citep{MASS-MK}.

The isoplanatic angle $\theta_0$ can  be computed from the restored TP
or  estimated directly  from the  scintillation  indices \citep{Rest}.
Temporal analysis of the light fluctuations in the smallest aperture A
permits  to estimate  the  time  constant $\tau_0$  by  the method  of
\citep{AO02}. This estimate is biased  because it does not include the
lowest 0.5\,km  of the atmosphere, but  it can be  corrected using the
DIMM data and known wind speed in the ground layer.

\subsection{DIMM: Differential Image Motion Monitor}
\label{sec:dimm}

 In a DIMM,  two circular portions of the  wavefront are isolated. Let
the diameter  of these sub-apertures  be $D$, their  separation (base)
$B$.    The  variance   of  the   differential  wave-front   tilts  in
longitudinal  (parallel to the base)  $\sigma^2_{l}$ and
transverse $\sigma^2_{t}$ directions is related to the Fried parameter
$r_0$ as \citep{DIMM,PASP02}
\begin{equation}
\sigma^2_{l,t} = K_{l,t} \; (\lambda/D)^2 \; (D/r_0)^{5/3} .
\label{eq:KDIMM}
\end{equation}
The {\em response coefficients} of  DIMM $K_{l,t}$ depend on the $B/D$
ratio and  on the  kind of  the tilt measured \citep{PASP02}. 

The wave-front tilts are estimated from the centroids of two images in
the  focal plane  of DIMM.   To reduce  the noise,  the  centroids are
calculated using only a sub-set of pixels selected either by setting a
threshold  well above  the background  noise spikes  or by  defining a
window around  the brightest pixel.  Both approaches  can be expressed
by a general formula
\begin{equation}
c_x = \sum_{i,j} w_{i,j} x_{i,j} I_{i,j}
/ I_0, \;\;\;
I_0 = \sum_{i,j} w_{i,j} I_{i,j},
\label{eq:centr}
\end{equation}
where $c_x$  is the  estimated centroid $x$-coordinate,  $I_{i,j}$ are
pixel intensities in arbitrary  units (most commonly in camera digital
counts,  ADUs),  $x_{i,j}$  are  their $x$-coordinates.   The  weights
$w_{i,j}$ equal one for selected pixels and zero otherwise.

By its principle, a DIMM is sensitive only to the phase distortions of
spatial  scales from  $D$ to  $B$.  Larger  scales  produce correlated
tilts, while smaller scales blur  the spots.  Usually both $D$ and $B$
are  in the  range where  the  Kolmogorov model  works well.   Optical
propagation has not  been considered by the standard  DIMM theory. The
propagation  reduces  the DIMM  response  by  two  effects --  partial
conversion of phase distortions into scintillation and deviations from
the  weak-perturbation  theory  (saturation).   Moreover,  even  small
optical  aberrations  can  significantly  bias the  DIMM  response  to
high-altitude turbulence. These effects are discussed in TK07.

Even  in  the  absence  of  atmospheric  image  motion,  the  measured
centroids fluctuate because  of the errors caused by  the photon noise
and detector  readout noise. The noise variance of each centroid is 
\begin{equation}
\sigma^2_c  = 
 \frac{1}{I_0^2} \sum_{i,j} (x_{i,j} - c_x)^2 (R^2 + I_{i,j}/G), 
\label{eq:sig3}
\end{equation}
where $R$  is readout noise in  ADU, $G$ is the  CCD camera conversion
factor (gain)  in $e^-$/ADU.  The sum in  (\ref{eq:sig3}) includes only
the pixels  used in the centroid  calculation.  It can  be computed in
advance if the  centroid window has a well-defined  size and the image
profile is known.  This is not  the case when a thresholding method is
used.   However, even  with thresholding  the centroid  noise  of each
individual spot can be  evaluated with (\ref{eq:sig3}) during centroid
computation.  Variations of the flux caused by scintillation or clouds
can be accounted for as well.

The differential  variance of the  wave-front tilts is  estimated from
the  finite number of  samples (frames)  $N$. The  lower limit  on the
relative statistical  error of the  measured variance $\sigma^2_{l,t}$
is  $\sqrt{ 2/N}$,  assuming  that the  samples  are independent.   To
measure the  seeing with a relative  error of 2\%, we  need to measure
the variance  with an error of $0.02  \times 5/3 = 0.033$,  hence $N >
1800$.  With typical 1\,min. accumulation time, a frame rate of 30\,Hz
or  larger is  required.   In  fact the  statistical  error of  seeing
measurements will  be larger  because frames are  partially correlated
and the seeing itself is non-stationary.

\section{MASS-DIMM instrument}
\label{sec:MASSDIMM}

\subsection{The instrument}

\begin{figure*}
\includegraphics[width=14cm]{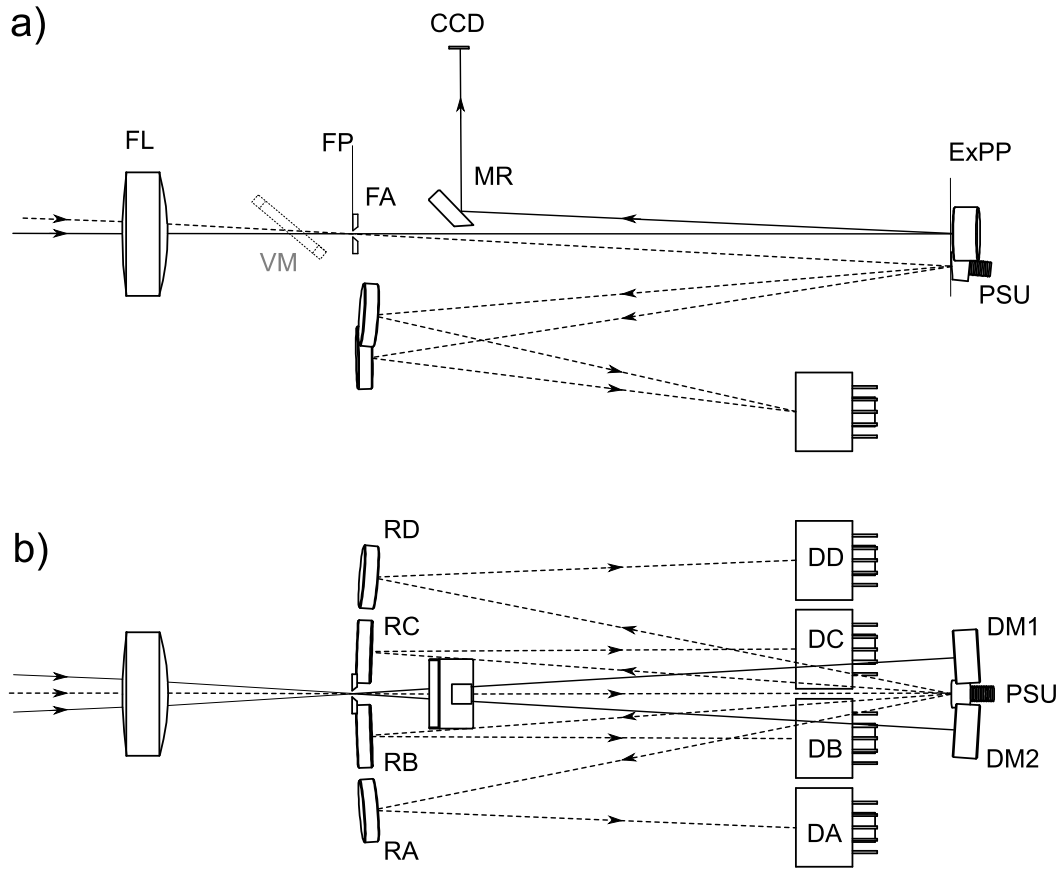} 
\caption{  Optical layout of  the MASS-DIMM  instrument. {\bf  a)} ---
side view,  {\bf b)} --- top  view. {\bf FP} ---  the instrument focal
plane, {\bf ExPP} --- the  plane of the exit pupil. Other designations
are explained in the text. 
\label{fig:MD} }
\end{figure*}

\begin{figure*}
\includegraphics[width=16cm]{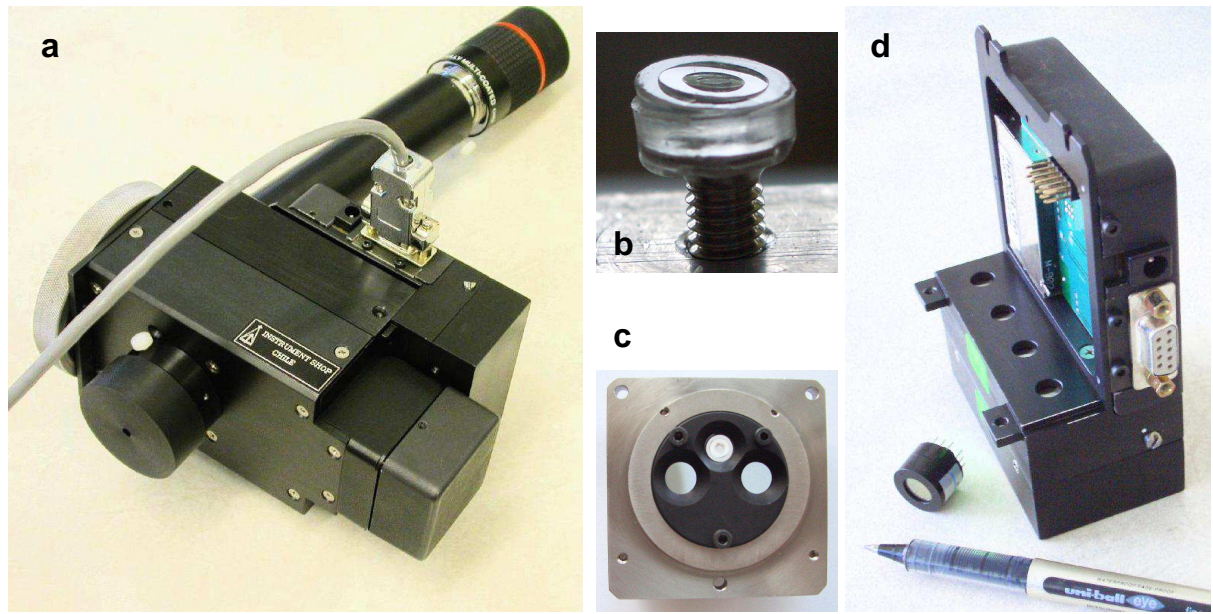} 
\caption{Details  of  the  MASS-DIMM  instrument.  (a)  The  assembled
instrument  with  a  protective  cup  installed  instead  of  the  CCD
camera. (b) Plastic segmentator replicated on top of the M3 screw. (c)
Pupil plate with a black mask. (d) Electronics module with 4 miniature
PMTs  and  photon  counters  inside,  one  PMT and  a  pen  are  shown
separately.
\label{fig:MDview}  }
\end{figure*}

A standard  DIMM instrument  uses only two  pieces of  the telescope's
aperture,  while  the  remaining  portion  could serve  for  the  MASS
channel. A study  of the optimum aperture size  \citep{Rest} has shown
that an  outer diameter of the  4 concentric MASS apertures  can be as
small as  8--9\,cm.  Such apertures  fit into the  annular un-obscured
zone  of   a  25--30  cm  Cassegrain   telescope  without  vignetting.
Combination  of MASS  and  DIMM  in a  single  instrument has  several
obvious advantages.  Only one telescope is required, with its pointing
and   tracking  controlled  by   centring  the   star  in   the  DIMM
channel. Both instruments  work on the same optical  path, thus sample
the same turbulent volume, so that non-stationarity does not introduce
any differential error.

The  optical   layout  of  the   MASS-DIMM  instrument  is   shown  in
Fig.~\ref{fig:MD}.  The MASS-DIMM instrument and its elements are shown
in  Fig.~\ref{fig:MDview}.  A weak  positive lens  (Fabry lens)  FL is
placed  in front  of the  telescope focus.   It slightly  shortens the
effective telescope focal  length and, at the same  time, forms a real
image  of the  pupil ExPP  at a  distance of  125~mm behind  the focal
plane.  A circular aperture FP  in the focal plane restricts the field
to a  typical diameter of 4$'$  in order to reduce  unwanted flux from
the sky and faint off-axis stars  in the MASS channel.  To ease manual
pointing, centring, and focusing, there  is a flip mirror VM in front
of the focal aperture followed by an eyepiece.

The size of the pupil  image ExPP depends on the optical magnification
factor $k_{\rm mag}$  of the telescope+lens system. For  a 25-cm Meade
telescope  and  125-mm  Fabry   lens,  the  magnification  is  $k_{\rm
mag}=14.5$, i.e. the pupil image has 17.4~mm diameter. The image falls
onto the  {\em pupil  plate} assembly that  actually splits  the light
between DIMM and MASS channels, described separately below.  The pupil
image on the  plate can be centred either by  moving the FL laterally
or  by tilting  the whole  MASS-DIMM  instrument with  respect to  the
telescope.

We assume that the entrance  pupil of the feeding telescope is located
on the top  end of the telescope tube, not on  the primary mirror.  In
two-mirror telescopes, the pupil  image is formed behind the secondary
mirror and  lies near the focal  plane of the primary  mirror.  The FL
projects it to  the ExPP plane.  The position and  focal length of the
FL must be chosen to  provide the required magnification $k_{\rm mag}$
in  this  plane.  By  changing  the FL,  we  can  adapt the  MASS-DIMM
instrument to  telescopes with slightly  different optical parameters.
The adaptation  possibility becomes  virtually un-limited if  a single
lens is replaced  by a two-lens system.  The  two-lens module performs
the same  function as a  single FL, i.e.   it produces the  real pupil
image with the required scale  $k_{\rm mag}$ and focuses the star onto
the field aperture at the same time.

\begin{table}
\center
\caption{Optical coupling of MASS-DIMM to some telescopes}
\label{tab:opt}
\begin{tabular}{l c c cccc}
\hline\hline
Telescope & $D$, & $F$, & $F_{\rm FL}$,& $k_{\rm mag}$ & $d_{\rm mask}$ & $b_{\rm mask}$ \\
          &  mm  & mm   &  mm  &       & mm            & mm \\[4pt]
\hline
Meade LX-200  & 252 & 2500 & 125          & 14.5 & 5.5   & 12.0  \\  
TMT-Halfmann  & 350 & 2800 & 140               & 15.6 & 6.4   & 15.5  \\ 
Meade RCX-400 & 305 & 2438 & 105\footnotemark[1]{}  & 16.9 & 5.5   & 12.5 \\  
Celestron C11 & 280 & 2800 & 120\footnotemark[2]{} & 16.0 & 5.5   & 12.1 \\[4pt]
 \hline\hline
\end{tabular}\\
{\small ${}^1$Two lenses: $F_1 = -100$, $F_2 = +75$} \\ 
{\small ${}^2$Two lenses: $F_1 = -150$, $F_2 = +75$} 
\end{table}

Table~\ref{tab:opt}  lists optical  parameters of  some telescopes
suitable for  MASS-DIMM. In each  case, we select matching  Fabry lens
and its distance from the focal plane and install a matched pupil mask
defining the  geometry of the  DIMM channel. The  magnification factor
$k_{\rm mag}$ depends on the axial position of the FL, which, in turn,
must  be adjusted  to  get a  sharp  pupil image  in  the ExPP  plane.
Instead of using nominal  $k_{\rm mag}$ from Table~\ref{tab:opt}, this
parameter must be carefully measured with an accuracy of at least $\pm
3$\% for each instrument-telescope combination.  When FL is a two-lens
combination, the  magnification can be tuned by  changing the distance
between the lenses  and their axial position.  It  also depends on the
small deviations  of the focal length  of the lenses  from its nominal
value.

\subsection{DIMM channel}

Two  spherical  mirrors  DM1  and DM2  (cf.   Fig.~\ref{fig:MD})  with
strictly identical  curvature radius  $R=135$~mm located in  the pupil
plane  reflect the light  back to  the focal  aperture, with  a slight
tilt.  The beams  are intercepted by a small flat  mirror MR tilted at
46$^\circ$ and directed to the CCD detector. The DIMM optics magnifies
the image in  the focal plane by 1.2 times.   Accurate focusing of the
stars in the DIMM channel is  critical, it is achieved by focusing the
telescope.

 An inter-changeable  mask in  the exit-pupil plane  has two  holes of
diameter $d_{\rm  mask}$ with centres separated by  $b_{\rm mask}$ and
actually  sets  the diameters  $D  =  d_{\rm  mask} k_{\rm  mag}$  and
separation $B  = b_{\rm mask} k_{\rm  mag}$ of the  DIMM apertures, as
projected  back to  the telescope  pupil. For  example,  $D=80$~mm and
$B=170$~mm  for Meade  LX-200.   The  tilts of  the  DIMM mirrors  are
fine-tuned  to  place each  spot  near the  centre  of  the CCD.   The
relative  position  of  the  spots  is  thus  fully  adjustable.  

The setup of  the CCD detector and spots depends  on the DIMM software
used.  In a typical DIMM with prism, the spots are separated along the
baseline and  the CCD orientation  is adjusted to make  this direction
parallel  to the  lines, so  that  the DIMM  software interprets  spot
motions  along  the  lines  as  longitudinal.   In  a  MASS-DIMM,  the
longitudinal  direction  is related  to  the  baseline  (hence to  the
mechanics of the instrument), not to the adjustable orientation of the
spots.  If the  same DIMM software is used, the lines  of the CCD must
be oriented parallel to the  baseline by rotating the detector package
relative to  the instrument.  Usually (but not  necessarily) the spots
are also placed  approximately along the lines by  tuning DM1 and DM2,
thus mimicking the configuration of a prism-based DIMM.

The  spherical  DIMM  mirrors  work  slightly off-axis  (the  tilt  is
$1.5^\circ$), but  the astigmatism  of this arrangement  is negligible
(Strehl ratio  $S>0.95$) because each of  the DIMM beams  is very slow,
typically $F/30$.  We document the optical quality of the DIMM channel
by taking slightly  defocused images of spots produced  by a simulated
point source at the entrance  of MASS-DIMM.  

\begin{table}
\center
\caption{CCD cameras for the DIMM channel}
\label{tab:ccd}
\begin{tabular}{l ll c c  l   }
\hline\hline
Product & CCD   & Format & Pixel, & $R$, & Interface  \\
name    & type  &        & $\mu$m & el.&  \\
\hline 
ST-5    & Frame+store &  320x240    & 10   & 20 & LPT port  \\
ST-7xME & Full  frame      & 765x510     &  9   & 15 & USB 1.1      \\
EC-650  & Inter-line       & 659x493     & 7.4  & 11 & IEEE 1394a \\          
 \hline\hline
\end{tabular}
\end{table}

Various  CCD cameras  can be  used as  detectors in  the  DIMM channel
(Table~\ref{tab:ccd}).  The requirement is to have small enough pixels
for  Nyquist sampling  of diffraction-limited  spots, $>2$  pixels per
$\lambda/D$, and  to enable short (5--10\,ms)  exposures. The distance
between the mask and the CCD is $l = 150$\,mm, so the spot size at the
CCD   is  $   l   \lambda   /  d_{\rm   mask}   =  17.7$\,$\mu$m   for
$\lambda=0.65$\,$\mu$m and $ d_{\rm mask} = 5.5$\,mm, calling for a
pixel size of  $\le 9$\,$\mu$m. 

Historically, we started with the  ST-5 camera from SBIG.  It contains
a frame-transfer  CCD with a storage  section, so that  short (down to
1\,ms) exposures  can be  taken.  The data  transfer to PC  limits the
acquisition rate to some 5 frames per second (FPS).  The DIMM software
suitable     for     ST-5    was     developed     for    the     CTIO
RoboDIMM\footnote{RoboDIMM:
http://www.ctio.noao.edu/telescopes/dimm/dimm.html}  and  works  under
Windows OS.

The ST-5 cameras are not produced any more and must be replaced by the
ST-7.   However, the  ST-7  has  no frame  storage  section and  hence
exposure by  mechanical shutter cannot  be shorter than  0.1\,s.  This
detector can  still be used  in a DIMM  in a special {\em  drift scan}
mode  where the  signal from  several lines  is vertically  binned and
these ``scans''  are read  out continuously at  a rate 5\,ms  per line
\citep{TMT-DIMM}.  Such scans permit to measure centroids only in one,
longitudinal direction.   In this mode, data  is acquired continuously
at  high  rate,  leading   to  low  statistical  errors  and  reliable
correction  of the exposure-time  bias.  The  software for  drift scan
DIMM operation was originally  developed by the Thirty Meter Telescope
(TMT) team and later re-written  at the Las Campanas Observatory (LCO)
by Ch.~Birk\footnote{LCO DIMM software:  http://www.ociw.edu/\~{}birk/CDIMM/}.  It works
only under Windows.

Recently, new CCD  detectors became available. A fast  CCD EC-650 from
{\em  Prosilica} is  compact (38x33x46\,mm),  robust, cheap,  with low
($<$2\,W) power consumption, and simple  to use. It enables very short
exposures and high data rate (90  FPS in full-frame, up to 300~FPS for
sub-region).  The DIMM  software for EC-650 (or any  other camera with
DCAM interface)  is developed by  V.~Kornilov under Linux\footnote{See
V.                           Kornilov,                         2006,\\
http://dragon.sai.msu.ru/mass/download/doc/dimm\_soft\_description.pdf}.
DIMM  can work  even with  cheap monochrome  8-bit cameras.   Tests of
colour  CCDs  used  in  some amateur  web-cam  DIMMs\footnote{DIMM  with
web-cam,                                                  C.~Cavadore,\\
http://astrosurf.com/cavadore/seeing/monitor\_DIMM/index.html}     have
shown that they have large  periodic errors caused by colour filters on
pixels and therefore should be avoided.

\subsection{MASS channel} 

The four concentric apertures are cut out from the exit pupil image by
a system of 4 tilted  annular mirrors called {\em segmentator} (PSU in
Fig.~\ref{fig:MD}).  First  segmentators were fabricated  by optically
polishing the ends  of the 4 matching bronze tubes cut  at an angle of
$8^\circ$ relative  to their  axis.  The tubes  are then turned  by an
angle  of   about  $30^\circ$  relative   to  each  other.    Now  the
segmentators are  fabricated by replicating a  pre-aligned master onto
acrylic plastic (Poly Methyl Metacrylate).  The optical quality of the
MASS channel is  not important because it only  measures the fluxes in
each  sub-aperture.  The  outer  diameter of  the largest  segmentator
mirror D is 5.5\,mm, the smallest mirror A is only 1\,mm.

  Upon reflection  from such  segmentator, the light  is split  into 4
distinct  beams, one  per aperture.   The beams  are intercepted  by 4
spherical mirrors (re-imagers) RA--RD  located around the focal plane,
and directed back  to 4 photo-multipliers (PMTs) DA--DD  arranged in a
linear  configuration  (Fig.~\ref{fig:MDview}).   The  images  of  the
segmentator apertures  are formed at  the photo-cathodes of  the PMTs.
The  electronics box  has a  manual  shutter for  protecting the  PMTs
during transport and installation.

Four miniature  PMTs R7400  from Hamamatsu are  used in MASS  as light
detectors.   They   are  located  in  the  box   together  with  their
pre-amplifiers,  discriminators  and  counting circuits.   The  photon
counter dead time  is only 16~ns.  The high  voltage (typically 800~V)
is  supplied  by  another  module  in  the  same  box.   This  modular
miniaturised   electronics   proved   to   be  very   convenient   and
reliable. The PMT  counts with 1\,ms sampling are  transmitted to a PC
computer working  under the Linux via RS-485  interface (4-way cable).
Sampling time  as short  as 0.25\,ms is  possible, although  only very
bright ($V < 1^m$) stars can  be observed in this mode.  A custom-made
adaptor converts serial  data to a form suitable  for input through PC
parallel (printer)  port.  The adaptor has  some buffering capability,
no  data is  lost.   Special  protocol provides  exchange  rate up  to
2~Mbit/sec.
The MASS electronics  and detectors can be self-tested  by an internal
light  source inside  the  instrument.  The  PMTs  are protected  from
over-light   by   automatically   cutting   high  voltage   when   the
photo-current exceeds some threshold.  Without such protection, costly
PMT replacements would have been inevitable.

\subsection{Operational algorithm}

\subsubsection{Supervisor}
MASS-DIMM instruments  typically operate  in robotic mode.   A program
called  {\em  Supervisor}  controls  the  telescope,  DIMM,  and  MASS
channels.    It  communicates   with  individual   components  through
sockets. Such architecture is very  flexible, it can run on one single
PC computer  or on several  computers. We need  at least 2 PCs  when a
Windows-based  CCD  software is  combined  with  the Linux-based  MASS
software.   The  DIMM software  developed  at  LCO  also functions  as
Supervisor.   Details  of  telescope  and  dome  control  vary  widely
depending on the  hardware and the site.  For  example, the Supervisor
may be connected  to a local meteo-station or to a control  room of a
large  telescope  to operate  MASS-DIMM  only  under suitable  weather
conditions.

The Supervisor selects a single bright star near zenith (air mass less
than 1.5) from a pre-defined  catalogue. The telescope is pointed to the
star and an image of the full field is taken with the CCD. If the star
is  not found  because of  bad initial  pointing or  clouds,  a spiral
search may be done. When the  star is finally acquired, it is centred
in the field and 1-minute integration in the DIMM and MASS channels is
started. Upon  the end of  the integration, the object  is re-centred
and the cycle  resumes until the air mass exceeds 1.5  and a change of
the star is required.  The sequence of suitable stars as a function of
sidereal time can be established in advance for any given site.

Apart from  the data acquisition, the  Supervisor ensures measurements
of  additional parameters  required  for correct  data processing  and
instrument control.   The sky background is  monitored periodically by
offsetting  the  telescope from  the  star.   The  MASS detectors  are
checked internally  to track their long-term variation.

\subsubsection{DIMM operation}
A  small  section  of  the  full  frame  containing  spots  is  imaged
repeatedly with short exposures.  In frame transfer CCDs, exposures of
$t_{\rm exp}$ and  $2 t_{\rm exp}$ are alternating  to account for the
exposure-time bias \citep{PASP02}. In a fast camera, when the temporal
gap between two adjacent exposures  is small, the binning method (as in
MASS)  can   be  used.    The  operation  in   the  drift   scan  mode
\citep{TMT-DIMM} is somewhat different and not discussed here.

Along with  centroids, the program calculates such  spot parameters as
integrated fluxes, Strehl ratios, ellipticity, etc.  Images spoiled by
telescope  wind shake (elliptical)  or clouds  can be  rejected.  Such
filtering  must   be  carefully  tuned   not to  disturb  the  seeing
statistics. The differential  variances in longitudinal and transverse
directions  are  calculated   for  the  remaining  (accepted)  images,
separately for  each exposure time. Estimates  of the ``longitudinal''
and ``transverse'' seeing  are then obtained using Eq.~\ref{eq:KDIMM},
averaged, and  corrected for  finite exposure time  \citep{PASP02} and
zenith distance.  Apart  from the seeing, the DIMM  data files contain
various  metrics  of  data  quality  such as  Strehl  ratios,  fluxes,
separation between spots, etc.

The Linux  software for the DIMM  channel operates in the  same way as
the MASS software.  Each {\em accumtime} (1\,min.) cycle is split into
{\em basetime} (2\,s) segments.   Such timing provides the calculation
of the  real accuracy of the output  data (including non-stationarity)
and periodic corrections of the ``stars box'' position.  The latter is
needed  to prevent  losing the  star  due to  telescope vibration  and
tracking.

\subsubsection{MASS operation}
Data acquisition  and processing  in the MASS  channel is done  by the
{\small  TURBINA} program  under Linux.   Each measurement  during the
{\em accumtime} (1\,min.)  cycle is  split into segments of the length
{\em  basetime}   (1\,s).   For  each   segment,  statistical  moments
(variances and  covariances) of the  photon counts are  calculated for
all 4 channels  and their combinations. The moments  are then averaged
over  {\em accumtime}.   Such two-stage  calculation filters  out slow
flux variations caused by unstable atmospheric transmission. Some MASS
data  obtained   through  thin  cirrus  cloud   pass  quality  control
(Sect.~\ref{sec:filter}) and  are valid.   It also allows  to estimate
the  real accuracy  of the  measured quantities  by  calculating their
variances.

After  each accumulation  time,  the scintillation  indices and  their
errors are  saved in the MASS  data file. This file  also contains the
turbulence profiles and other  atmospheric parameters derived from the
indices  and  various auxiliary  quantities  useful  for data  quality
analysis.  The  statistical moments for each {\em  basetime} are saved
in a  separate file  and used later  for re-processing,  if necessary.
The  moments are  raw data  independent  of any  model and  instrument
parameters.  Also, some auxiliary operation modes, such as background
estimation, detector and statistics tests are realised.

\section{Use of MASS-DIMM}
\label{sec:use}

Like any instrument, MASS-DIMM can provide inaccurate or wrong data if
not used correctly. Here we outline the necessary procedures.

\subsection{Setting MASS parameters}

MASS is  essentially a  fast photometer, its  signal is  interpreted in
terms  of seeing  using  only theoretical  WFs. However,  instrumental
parameters  have  to  be  specified  accurately to  ensure  a  correct
calculation of the indices and the WFs. 

\subsubsection{High voltage and discriminator thresholds}
For   each   set  of   MASS   detectors   and  electronics,   counting
characteristics as  a function of  the high voltage  and discriminator
thresholds  are  measured during  assembly  and testing.   Recommended
thresholds and high  voltage values for each device  are determined to
ensure the  lowest noise.   Needless to say  that when  the electronic
modules  or detectors are  replaced, the  configuration files  must be
updated, too.

\subsubsection{Non-Poisson parameter}
The  parameter   $p  \approx  1$  quantifies  the   deviation  of  the
photon-counting  statistics from  the Poisson  law  \citep{Rest}.  The
deviation is  only few  percent, but  a wrong $p$  value leads  to the
over- or under-estimation  of the photon noise and hence  to a bias in
the computed indices, especially those involving small apertures A and
B.  The effect becomes important under good seeing when typical values
of $s^2_{AB}$  are $\approx 0.02$ only.   This means that  if the star
flux  is less than 50  pulses per  integration, over  half of  the measured
signal variance  is due to the  shot noise and  its correct estimation
becomes critical.   It is preferable  to select brighter  target stars
for  reducing the  dependence of  results  on the  correct $p$  value.
Typically, a $V  \approx 2^m$ star produces about 50  counts per ms in
aperture A, 100 in B, 300 in C, and 500 in D.

The non-Poisson  parameters must be  regularly re-measured  using internal
light  source  or  sky  background  to track  possible  ageing  of  the
detectors or drifts of  the high voltage and discriminator thresholds.
The statistical error of the $p$-parameter estimate is
\begin{equation}
\epsilon_{p}^2 = \frac{2}{N} \left( 1+ \frac{1}{F} \right),
\end{equation}
where $N$ is a total number  of micro-exposures, $F$ is the mean count
per  micro-exposure. To achieve  the relative  precision of  $p$ about
0.5\%,  one needs  the  accumulation time  of  more than  100~s at  $F
> 10$. The parameter $p$  depends on the temperature and settles
slowly after  the HV is  turned on.  So,  we recommend to  measure $p$
during the night time, approximately 1 hour after switching on the HV.
Yet another way  to check the correct setting  of $p$-parameters is to
do  a {\em  statistical test}  when the  light source  inside  MASS is
modulated.    The  differential   indices  in   some   aperture  pairs
(especially  AB)  will  systematically   differ  from  zero  when  the
$p$-parameters are incorrect.

\subsubsection{Dead time}
The non-linearity parameter (dead time)  $\tau$ plays a role only when
the  target star  is brighter  than  $V=1^m$.  It  affects mainly  the
scintillation indices in  apertures D and C.  The  effect of incorrect
$\tau$ setting is  multiplicative and small, it changes  the TP by few
percent, at most.

\subsubsection{Magnification}
The  size   of  the  MASS   apertures  $D$  depends  on   the  optical
magnification factor  $k_{\rm mag}$. Approximate  analytics shows that
$z  \sim D^2$  and the  turbulence integral  (Eq.~\ref{eq:J})  $J \sim
D^{5/3}$, so the error of  3\% in $k_{\rm mag}$ translates to $\approx
5\%$  in  turbulence  intensity  and  6\% in  layers  altitudes.   Our
numerical  calculations confirm  this estimate.   Since  $k_{\rm mag}$
depends on the alignment of both feeding telescope and MASS, it cannot
be determined  from the optical  design parameters but rather  must be
measured on the real device.  To  do so, we either measure the size of
the exit  pupil image  or send  a bright light  beam back  through the
instrument and measure its footprint on the entrance aperture.

\subsubsection{Vignetting}
Vignetting  of  the  MASS  apertures  must be  avoided  at  all  cost.
Modification of the aperture shape by vignetting and other causes such
as dirty optics influences the  WFs. This effect is controlled to some
extent by monitoring  the flux ratio between MASS  channels which must
remain  constant to  within few  percent.  We  tend to  select maximum
projected  size  of  MASS  apertures  (i.e.   maximum  $k_{\rm  mag}$)
permitted by the telescope's aperture to collect more light and to get
a  more robust  profile  restoration.  As  a  downside, tolerance  for
mis-alignment  becomes  small,  hence   it  is  essential  to  control
vignetting. When MASS is  installed at the telescope and the Fabry
lens is aligned, we detach the electronics module and examine visually
the  image of  the  entrance pupil  in  the D  channel  for traces  of
vignetting.    Vignetting must  be  checked  each  time after  the
instrument was  removed from  the telescope or  after any  other major
changes.

\subsubsection{Spectral response}

\begin{figure}
\includegraphics[angle=-90,width=8cm]{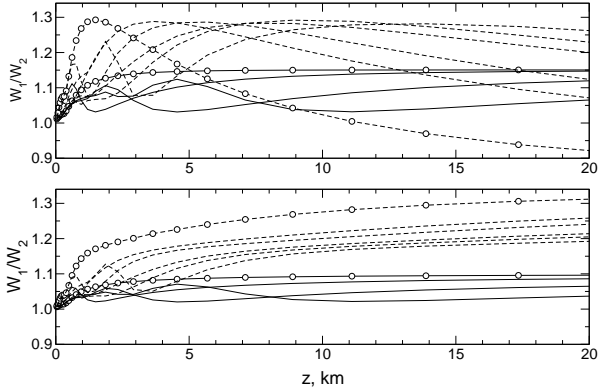} 
\caption{ Top --- the ratio of WFs for the spectral response of a MASS
 with a yellow  cutoff filter to those without  filter. Bottom --- the
 ratio of WFs for stars of  spectral types B0V and G0III. Solid curves
 represent normal indices, dashed curves -- differential ones, $s^2_A$
 and $s^2_{AB}$ are marked by circles.
\label{fig:WFR}  }
\end{figure}

The spectral  response of the MASS  instrument $F(\lambda)$ influences
the  WFs \citep{Poly}.  The  response curves are determined  from the
known characteristics  of the  MASS components (detector,  optics) and
multiplied by the known spectral  energy distribution in the source (a
star of  known spectral type)  to calculate $W_k(z)$.  We  checked the
correspondence of  the assumed $F(\lambda)$ with the  actual fluxes of
stars  of  different  colours  and  found disagreements  in  some  MASS
instruments.\footnote{V. Kornilov, 2006,  The verification of the MASS
spectral                         response.                          \\
http://www.ctio.noao.edu/\~{}atokovin/profiler/mass\_spectral\_band\_eng.pdf}
Figure~\ref{fig:WFR}  plots the ratio  of the  WFs calculated  for two
different spectral responses and different star spectra.  In the worst
case, wrong  definition of the  spectral response or  stellar spectral
type biases the  WFs by as much  as 30\%, with a larger  effect on the
differential indices.

Originally, MASS devices had yellow filters blocking the UV light with
$\lambda <  450$\,nm.  The filters  were removed to increase  the flux
because the PMTs with bi-alkali photo-cathode have maximum sensitivity
near 400~nm. However, the spectral response in the UV is now dependent
on the reflectivity curve of  telescope mirror coatings.  We advise to
 check the spectral response  of each MASS device periodically (1 or
2 times per year) by  measuring fluxes from photometric standard stars
of different colours and interpreting  the results as explained in the
above-cited document.

In short,  the results of MASS  are accurate only  when its parameters
are specified  correctly. Several checks to evaluate  the data quality
{\em a posteriori}  have been developed,   but the best way is to control
instrument parameters during  measurements.

\subsection{DIMM parameters}
\label{sec:DIMM}

\subsubsection{Pixel scale}
The  angular size  of the  CCD pixel  must be  known to  transform the
measured   centroid   variance   $\sigma^2_{t,l}$   from   pixels   to
radians. The pixel  scale must be measured by  taking images of double
stars  with   known  separation.  We   cannot  rely  on   the  nominal
(calculated)  pixel  scale because  all  individual  telescopes are  a
little different  and, moreover, the  scale depends on the  Fabry lens
position, detector  position, and  telescope focus.  The  double stars
HR7141/2  = $\theta$  Ser  ($\rho=22\farcs3$), HR5984/5  = $\beta^1$  Sco
($\rho=$14\farcs0), HR8895  ($\rho=26\farcs5$), HR1886/7 ($\rho=36\farcs2$) are
suitable for these determinations.

\begin{figure}
\includegraphics[width=8cm]{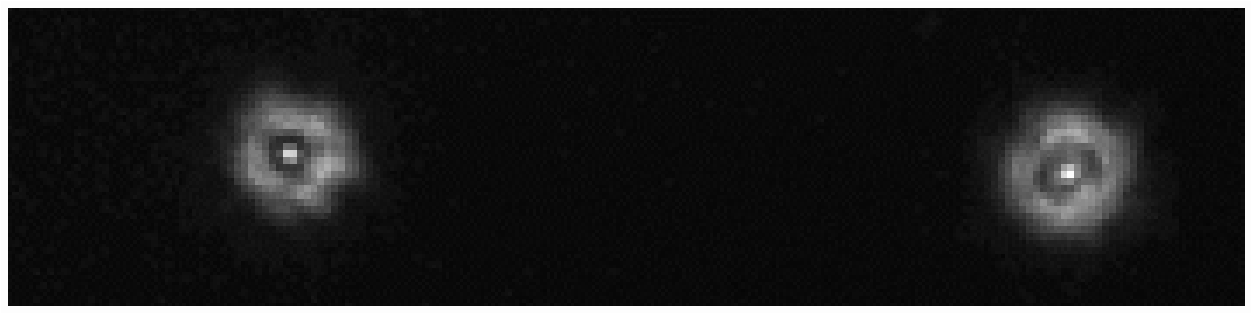} 
\includegraphics[width=8cm]{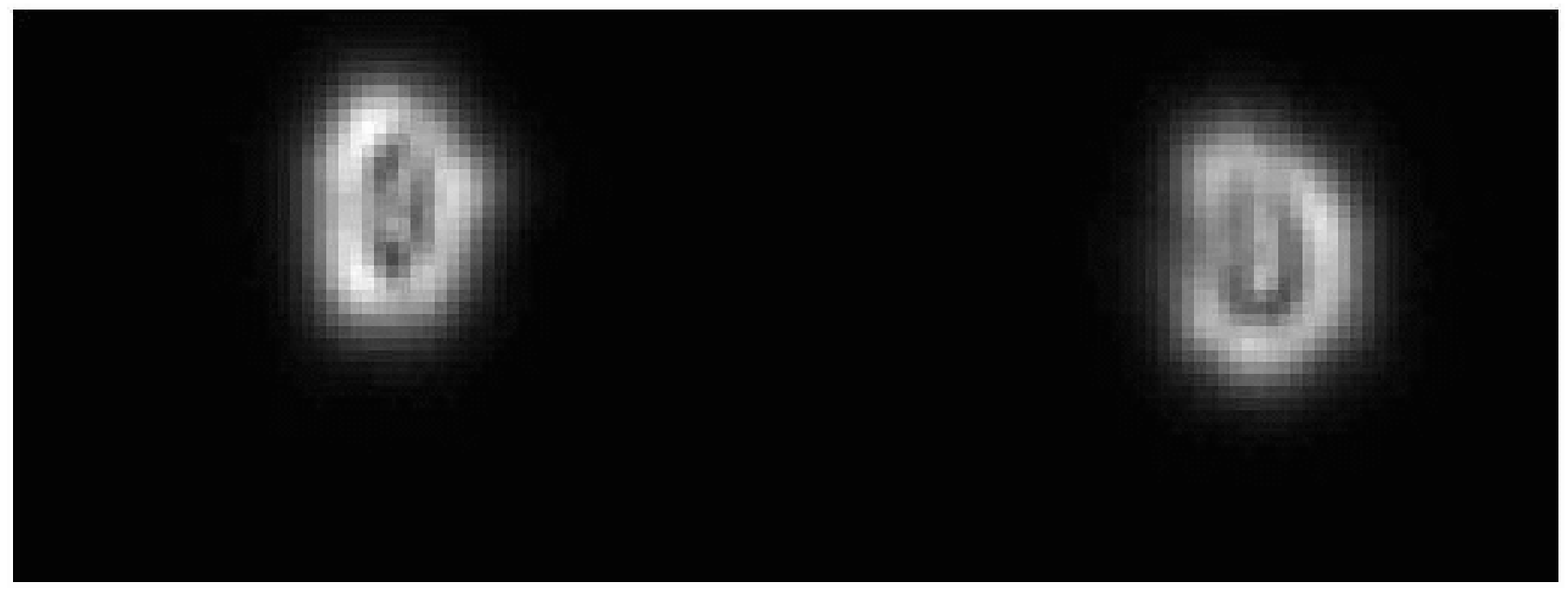} 
\caption{Pairs of defocused spots  in the DIMM channel: nearly perfect
  spots seen with a point-source simulator in the laboratory (top) and
  astigmatic  (about  60\,nm rms)  spots  produced  by  a star  and  a
  telescope with  spherical aberration (bottom) which  correspond to a
  Strehl ratio of $\sim$0.6.
\label{fig:defoc} }
\end{figure}

\subsubsection{Optical quality}
Until  recently,  the  optical  quality  of the  spots  has  not  been
identified  as  a  major  source  of systematic  errors  in  DIMM.   A
significant  bias in  the seeing  is  actually caused  by the  complex
interaction between aberrations and scintillation, as evidenced by the
study of  TK07 and some experiments  \citep{TMT-DIMM}.  By maintaining
nearly-diffraction image quality, $S > 0.6$, we avoid strong bias even
for high-altitude turbulent layers.

When  the  MASS-DIMM  system  is  set, the  optical  quality  must  be
controlled by  taking long-exposure  ($\ge 30$\,s) images  of slightly
defocused  spots  of  some  faint  star under  good  seeing  and  good
tracking.  Atmospheric distortions are averaged out, while aberrations
manifest    themselves    as    distortions    of    the    ``donuts''
(Fig.~\ref{fig:defoc}).  The aberrations can be quantified by the {\em
donut} method of \citet{Donut}.  This check must be repeated regularly
because the  telescope alignment is never very  stable. Astigmatism in
the direction of the baseline is acceptable in the drift-scan mode.

The defocus  caused by temperature changes  and mechanical instability
in  the  telescope  is  usually   a  major  contributor  to  the  spot
degradation.  The MASS-DIMM instrument has no internal focusing stage,
while the focusing of amateur  telescopes is manual.  There is no free
clearance in the  standard Meade fork mount for  using motorised focus
adaptors with  MASS-DIMM and LX-200.   In this respect,  Meade RCX-400
with  motorised  focus  is  better\footnote{Unfortunately,  all  Meade
RCX-400 telescopes used to date with MASS-DIMMs have failed due to the
manufacturer's error in their control software}.

The separation  of the spots  $\Delta x$ changes with  telescope focus
and  can serve to  stabilise it  when the  motorised focus  control is
available.  Of course, the separation also depends on the alignment of
the MD1  and MD2 mirrors, but  our experience shows that  it is stable
enough.  For a given MASS-DIMM system, the plot of Strehl ratio versus
$\Delta x$ shows a broad maximum corresponding to the optimum focus.

The  degradation of  the Strehl  ratio $S$  caused by  defocus  can be
quantified  by  noting that  the  corresponding  rms phase  aberration
$\sigma_{\varphi}$  is  related to  the  change  in  the angular  spot
separation $A$, 
\begin{equation}
S_{\rm defoc.} \approx S_0 \exp[ - \sigma_{\varphi}^2] = 
S_0 \exp \left[ - 
\frac{\pi^2}{48} \left( \frac{A D}{\lambda}\right)^2 
 \left( \frac{D}{B} \right)^2 \right] .
\label{eq:Strehl}
\end{equation}
The $A$  can be  converted from  radians to pixels  if divided  by the
angular pixel scale.  If we  require $S/S_0 > 0.8$, this translates to
$A  <  1.04  (\lambda/D)  (B/D)$,  or about  2.6  arcsec  for  typical
parameters  $D =  0.1$\,m, $B/D=2.5$,  $\lambda =  0.5$\,$\mu$m.  This
gives an idea of the acceptable tolerance on the spot separation.  The
defocus  tolerance  is  proportional  to $D^{-2}$,  favouring  smaller
apertures, so typical $\sim$8\,cm DIMM apertures are a good compromise
between sensitivity and robustness.

\subsubsection{Centroid noise}
It follows from Eq.~\ref{eq:sig3} that for a correct estimation of the
noise bias, two CCD camera parameters must be known: the readout noise
$R$ and the camera gain $G$.   Both are measured in a standard way, by
computing  signal  variance  at  different illumination  levels.   The
readout  noise  can  be  also   estimated  from  the  data  frames  by
calculating the  background variance.   Typically, the noise  is small
and its  subtraction or not does  not matter.  However,  the noise can
significantly bias DIMM results under good seeing or for faint stars.
We strongly recommend to calculate the noise variance of {\em each spot} 
in real time and subtract the noise contribution of both spots from
the measured differential variance. 

\subsubsection{CCD pattern noise}

 Many front-illuminated  CCD detectors have regular  variations of the
sensitivity  over their  pixels.   Depending  on a  type  of CCD,  the
sensitivity  variations may  be in  the horizontal  (as  in inter-line
CCDs) or vertical  (as in frame transfer CCDs)  directions.  When size
of  a  star  image  is   close  to  Nyquist  limit,  such  intra-pixel
sensitivity  variations cause  errors in  the centroids,  depending on
spot position.   Thus, the common motion  of the image pair  in a DIMM
can  produce   the  variation  of  the  image   separation.   In  real
measurements,  common  motion is  normally  present  due to  telescope
vibrations, tracking errors, and turbulence.

Naturally, if the CCD has intra-pixel sensitivity variations along the
lines (horizontal), the longitudinal variance will be affected only by
the horizontal motion.   In a chessboard-like pattern (as in colour
CCDs), both  horizontal and vertical motions of the spots will cause
errors in both longitudinal and transverse directions. 

\begin{figure}
\includegraphics[angle=-90,width=8cm]{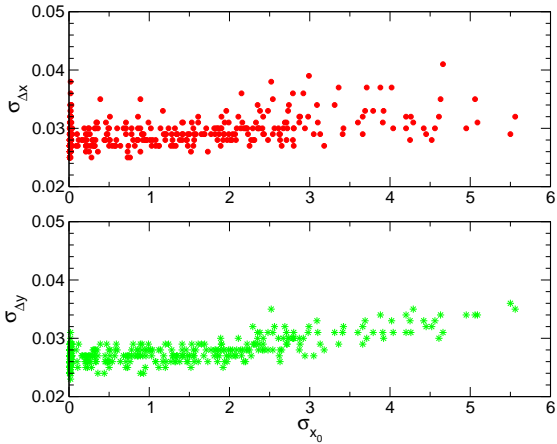}
\caption{ Dependence of measured differential variance in longitudinal
  ($\sigma_{\Delta  x}$,  top)  and transverse  ($\sigma_{\Delta  y}$,
  bottom) directions  on the r.m.s.  amplitude of  the artificial star
  motion  (both  in  pixels)   for  the  Prosilica  CCD  camera.   The
  calculated  centroid noise  is about  0.027 pixels.  \label{fig:mod} }
\end{figure}

To be sure  that this effect is negligible for a  chosen CCD camera, a
simple test must be done with a star simulator. We move the arfificial
star by  few pixels  during the seeing  measurements.  If there  is no
modulation, the  total variance will correspond to  the centroid noise
(Eq.~\ref{eq:sig3}).   Fig.~\ref{fig:mod}  shows  the result  of  such
experiment with the Prosilica EC650  camera, demonstrating clearly the
abscence of the pattern noise.  A weak increase of the differential
variance in  Fig.~\ref{fig:mod} is likely explained by the large-scale
sensitivity variation of the CCD and smearing of spots.  

\subsection{Data quality control}

A working site  monitor with a MASS-DIMM inevitably  produces some bad
data,  to  be  discarded  {\em  a  posteriori}  by  applying  suitable
criteria,  {\em filters}.   Any such  filtering must  not  distort the
statistical distribution of  the measured atmospheric parameters. When
the  system works  correctly,  only a small fraction  of  the data  is
rejected.

\subsubsection{DIMM data filters}
The  Strehl  ratio $S$  is  a good  measure  of  the optical  quality.
However, even in  a perfect DIMM $S$ is reduced  under bad seeing when
$r_0 <  D$. Hence,  by setting a  fixed threshold $S_0$  and rejecting
data with $S  < S_0$ we bias the seeing  statistics.  A more elaborate
seeing-dependent  threshold should  be used  instead \citep{TMT-DIMM}.
Alternatively, a  relation between $S$ and spot  separation $\Delta x$
is established empirically and the  data are filtered by condition $ |
\Delta x - \Delta x_0 | <  A$, where $ \Delta x_0$ is the optimum spot
separation  and  $A$  is  the  acceptable  half-range  evaluated  from
(\ref{eq:Strehl}).

\subsubsection{MASS data filters}
\label{sec:filter}
Various parameters stored in the MASS data files can be used to select
valid  data.    The  rejection   thresholds  are  specific   for  each
instrument,  so  the values  given  in  Table~\ref{tab:filt} are  only
indicative.

\begin{table*}
\center
\caption{MASS data filters}
\label{tab:filt}
\begin{tabular}{l l l   }
\hline\hline
Parameter & Rejection & Rejection reason \\
\hline 
Flux in D-aperture $F_D$  & $F_D < 100$ & Faint star or clouds \\
Flux error $\delta F_D$   &  $\delta F_D > 0.01$ & Cirrus clouds, bad guiding \\
Model error $\chi^2$      & $\chi^2 > 100$ &  Bad profile restoration \\
Relative background $B_D/F_D$ &  $B_D/F_D > 0.03 $ & Bright sky or star in the aperture  \\
\hline\hline
\end{tabular}
\end{table*}

In addition  to criteria on individual measurements,  the sanity check
on the  data set as  a whole is  done by calculating  average relative
residuals between scintillation indices and their model (the residuals
are saved).  Average residual exceeding 5\% on any of the indices is a
strong indication  of some problem  such as wrong  parameter settings,
scattered light or vignetting. The vignetting is also monitored by the
average flux  ratio $F_C/F_D$ which  increases when the aperture  D is
vignetted.   The ratio  depends slightly  on the  colours of  the stars
because individual PMTs have slightly different spectral response.

\subsection{Reprocessing}

MASS data compromised  by the wrong setting of  some parameters can be
recovered by  re-processing. For example,  we can study  the long-term
variation of the $p$-parameters and then apply their correct values to
the old data.  A wrong setting  of the computer clock (hence wrong air
mass  calculation)  can  be  corrected,  too.   A  set  of  tools  for
re-processing   is  provided  together   with  the  {\small TURBINA}  program.
Individual statistical moments with 1-s integration are used in the
re-processing and  must be archived together with the main MASS data files.

Starting from March 2006, a new algorithm for correcting the effect of
strong  scintillation (over-shoots) is  implemented in  {\small TURBINA}.
All  previous data  must be  re-processed with  this new  algorithm to
correct   over-shoots    and   TP   distortion    caused   by   strong
scintillation. The  correction method and an  example of re-processing
are given in TK07.

\section{Examples and applications}
\label{sec:examples}

\subsection{Data examples}

\begin{figure}
\includegraphics[angle=270,width=8cm]{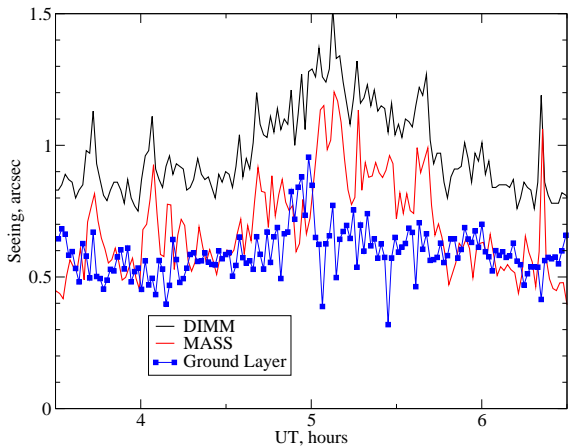} 
\caption{Data from the MASS-DIMM at Cerro Tololo, February 7/8 2007.
The total seeing measured by DIMM, free-atmosphere seeing from MASS
and the ground-layer seeing derived by subtraction is plotted. 
\label{fig:Feb07} }
\end{figure}

Figure~\ref{fig:Feb07}  shows  a plot  of  the  seeing  measured by  a
MASS-DIMM site monitor  at Cerro Tololo on the  night of February 7/8,
2007. This  night  was  chosen  because  a  significant  part  of  the
turbulence was  located at an altitude  of $\sim 4$\,km  and sensed by
both  instruments.  The  spikes  of seeing  are perfectly  correlated.
When  the ground-layer  (GL) seeing  is  calculated as  $J_{\rm GL}  =
J_{\rm  DIMM} -  J_{\rm MASS}$  (MASS corrected  for  over-shoots, $J$
defined  by  Eq.~\ref{eq:J}),  these  spikes  disappear.   This  is  a
convincing demonstration of the fact that both instruments measure the
seeing on the same, absolute scale.  Both DIMM and MASS are calibrated
independently, without any mutual adjustments or corrections.

The subtraction procedure illustrated in Fig.~\ref{fig:Feb07} does not
work   when  the   scintillation   becomes  strong,   $s^2_A  >   0.7$
(cf.  Eq.~\ref{eq:sig_I}).  Under  these  conditions,  MASS  typically
over-estimates the seeing even after over-shoot correction, while DIMM
under-shoots.  Both instruments thus  lose their accuracy under strong
scintillation.

\begin{figure}
\includegraphics[angle=270,width=8cm]{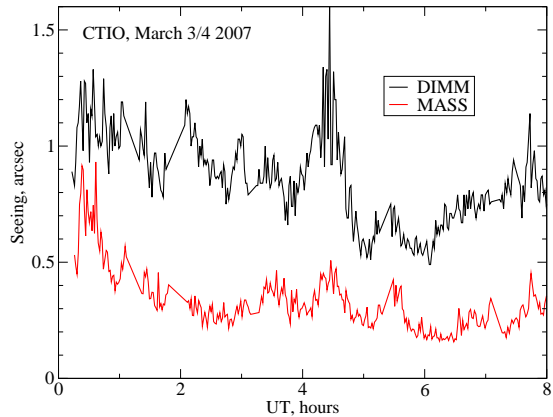} 
\includegraphics[angle=270,width=8cm]{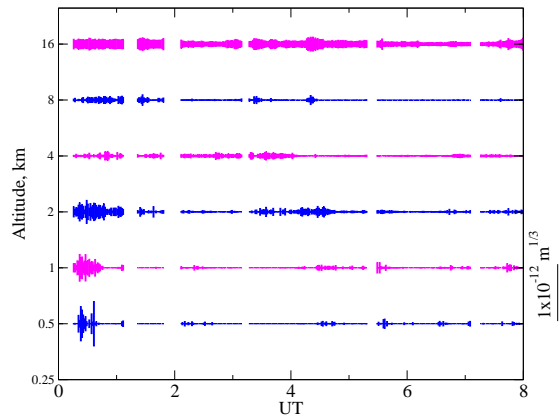} 
\caption{Data from the MASS-DIMM at Cerro Tololo, March 3/4 2007.  The
total seeing measured by DIMM was dominated by the ground layer, while
the  free-atmosphere  seeing  was  very good  (top).  The  turbulence
profiles (bottom)  show that,  except for the  beginning of  the night,
most turbulence was located in the highest 16-km layer.
\label{fig:Mar03} }
\end{figure}

Half or  more of the total turbulence  integral is often produced
by  the GL.  The  GL contribution  can be  estimated then  reliably by
subtraction. On some nights, the seeing in the free atmosphere is good
and stable, between 0\farcs2 and 0\farcs3 (Fig.~\ref{fig:Mar03}).  These
conditions,   encountered   infrequently    but   regularly   at   all
observatories where MASS-DIMM  operates, resemble Antarctic sites
\citep[e.g.][]{DomeC}.  If
ground-layer  turbulence were  compensated by  AO, an  excellent image
quality over wide field could  be achieved on these calm nights.  Such
special conditions cannot  be recognised  only with a  DIMM because the
total seeing  on these calm nights  is always dominated by  the GL and
not at all exceptional. Such  calm periods can last for several nights
\citep[e.g.][]{Tok03}. At Cerro  Pach\'on, the free-atmosphere seeing is
better than 0\farcs29 during 25\% of clear night time \citep{TT06}.

\subsection{Applications}

\subsubsection{Site testing}
A  possibility to  monitor turbulence  in the  free atmosphere  with a
small and  robust instrument  such as  MASS is of  high value  to site
testing. MASS is  not sensitive to small pointing  errors and can work
even from a heated room through a window, as demonstrated by the first
night-time seeing measurements in Antarctica \citep{DomeC}.

The development of the  MASS-DIMM instrument has been primarily driven
by  the need to  characterise new  and existing  sites for  the Thirty
Meter  Telescope (TMT) project  \citep{TMT-DIMM}.  Six  MASS-DIMMs were
deployed  together  with  robotic  site-testing telescopes  and  other
equipment.   The data  collection is  being continued.   Prior  to the
deployment, the equipment has been extensively tested at Cerro Tololo.
A European  program of site selection  for a large  telescope will use
four MASS-DIMMs. The program of Las Campanas site characterisation for
the 20-m Giant Magellan Telescope uses one MASS-DIMM.

The  advantage   of  a  MASS-DIMM   is  obvious,  as  it   permits  to
differentiate atmospheric regions where  seeing is degraded, whereas a
simple DIMM  measures only integrated  seeing. Significant differences
between sites with  comparable seeing caused by the  GL turbulence now
become  apparent.  Moreover, measurements  of $\theta_0$  and $\tau_0$
provided by MASS are essential for selecting sites better suited for AO.

\subsubsection{Site monitoring}
First systematic  measurements of  TP at Cerro  Tololo have  been done
with separate MASS and DIMM monitors \citep{Tok03}. Combined MASS-DIMM
site monitors  based on  25-cm Meade telescopes  work in  robotic mode
both at  the Cerro  Tololo and at  Cerro Pach\'on  observatories since
2004.  The   data  from   MASS  are   available  publicly\footnote{See
http://139.229.11.21/ },  the data from  DIMMs -- upon  request. These
openly available  data are used  in some studies  \citep{Kenyon}.  The
statistical analysis of the MASS-DIMM  data for Cerro Pach\'on shows a
good agreement with previous data sets  and leads to a new model of TP
at this observatory \citep{TT06}.

High photometric stability of the MASS detectors permits to study the
atmospheric  transparency and  its  variations (in  the MASS  spectral
band) and to obtain photometric characterisation of a site. 

\subsubsection{Support of adaptive optics}
Statistics  of TP and  $\tau_0$ delivered  by MASS-DIMMs  gives useful
input  for predicting performance  of current  and future  AO systems.
\citet{Britton06} compared the actual anisoplanatism in AO images with
calculations based  on simultaneous TPs from MASS-DIMM  and found that
the latter provide  a very good diagnostic. A  large collection of TPs
at Cerro Pach\'on  served to evaluate the gain  from a Ground-Layer AO
(GLAO) in  statistical sense \citep{GLAO}.   A similar GLAO  study for
Gemini \citep{Gemini-GLAO}  is based on  the TP model, while  the GLAO
study  for  the  Magellan  telescope  used  real  TPs  from  MASS-DIMM
\citep{Athey06}.

Real-time  monitoring  of  TP  with  MASS-DIMM opens  a  new  exciting
possibility  to  schedule  critical  AO observations  flexibly,  taking
advantage  of the  best  available conditions.   These conditions  are
comparable  to the  calm atmosphere  over Antarctic  plateau,  but are
encountered at  existing observatories with 10-m  class telescopes and
an overall seeing better than in Antarctica.  Potential improvement of
AO capabilities  (more uniform correction in a  wider field, operation
at visible  wavelengths) is too important  to miss and  gives a strong
incentive for flexible AO scheduling.

To date, a total of  22 MASS-DIMM instruments have been fabricated and
delivered  to  several  observatories  and programs.   There  are  all
reasons  to believe  that these  instruments will  make  a significant
contribution to  our understanding of atmospheric  turbulence and will
extend the  potential of  new methods such  as AO  and interferometry.
MASS-DIMM  becomes   a  standard  instrument  for   site  testing  and
monitoring.

\section*{Acknowledgements}

 The  development  of the  MASS-DIMM  instrument  and  the methods  of
getting   accurate  seeing  measurements   has  been   stimulated  and
encouraged  by  many people  and  organisations.   We acknowledge  the
support  from  NOAO  and  ESO  in building  and  testing  the  initial
prototypes and final instruments.  Several testing campaigns have been
done at CTIO in the period 2002-2004.
The software of  the MASS instrument has been  developed and supported
by the team  at the Sternberg Institute of  the Moscow University. The
mechanical parts were produced at  the CTIO Workshop.  We are indebted
to  the CTIO  ``sites  group'' (E.~Bustos,  J.~Seguel, D.~Walker)  for
maintaining MASS-DIMM  site monitors in a  working and well-calibrated
condition. The comments of anonymous Referee helped to improve the
article.


\bsp

\label{lastpage}


\begin{thebibliography}{99}


\bibitem[\protect\citeauthoryear{Anderson et al.}{2006}]{Gemini-GLAO}
Andersen D.R., Stoesz J., Morris S.,
Lloyd-Hart M., Crampton D., Butterly T., Ellerbroek B., Jolissaint L. et al.,
2006, PASP, 118, 1574


\bibitem[\protect\citeauthoryear{Athey et al.}{2006}]{Athey06}
Athey A., Shectman S., Phillips M., Thomas-Osip J., 
	 2006, 
in Ellerbroeck B.L.,  Bonaccini C.D., eds., Advances in Adaptive Optics II, 
Proc. SPIE, 6272, 627217

\bibitem[\protect\citeauthoryear{Britton}{2006}]{Britton06}
Britton M. C., 2006, PASP, 118,  885


\bibitem[\protect\citeauthoryear{Habib et al.}{2006}]{Habib}
Habib A., Vernin J. Benkhaldoun Z.,  Lanteri H.,
2006, MNRAS, 368, 1456


\bibitem[\protect\citeauthoryear{Hardy}{1998}]{Hardy}
Hardy J.W. 
1998, Adaptive Optics for Astronomical Telescopes.
Oxford Univ. Press, Oxford

\bibitem[\protect\citeauthoryear{Kenyon et al.}{2006}]{Kenyon}
Kenyon S., Lawrence J.S., Ashley M.C.B., Storey J.W.V., Tokovinin A., Fossat E.,
2006, PASP, 118, 924

\bibitem[\protect\citeauthoryear{Kornilov et al.}{2003}]{MASS}
Kornilov  V., Tokovinin  A.,  Voziakova O., 
Zaitsev A., Shatsky N., Potanin S., Sarazin M., 
2003, Proc. SPIE, 4839, 837

\bibitem[\protect\citeauthoryear{Lawrence et al.}{2004}]{DomeC}
Lawrence J.S., Ashley M.C.B., Tokovinin A.,  Travouillon T., 2004,
Nature, 431, 278


\bibitem[\protect\citeauthoryear{Roddier}{1981}]{Roddier81}
Roddier F.  1981, in Wolf E., ed. Progress in Optics, Vol. 19. 
North-Holland, Amsterdam, p. 281


\bibitem[\protect\citeauthoryear{Sarazin    \&   Roddier}{1990}]{DIMM}
Sarazin M., Roddier F.,  1990, A\&A, 227, 294


\bibitem[\protect\citeauthoryear{Tatarskii}{1961}]{Tatarskii}
Tatarskii V.I., 1961, Wave Propagation in a Turbulent Medium.
Dover Press, New York

\bibitem[\protect\citeauthoryear{Tokovinin}{2002a}]{PASP02}
Tokovinin A., 2002a, PASP, 114, 1156


\bibitem[\protect\citeauthoryear{Tokovinin}{2002b}]{AO02}
Tokovinin A., 2002b,  Appl. Opt.,  41,  957 

\bibitem[\protect\citeauthoryear{Tokovinin}{2003}]{Poly}
Tokovinin A., 2003,   J. Opt.  Soc. Am. A, 20,  686



\bibitem[\protect\citeauthoryear{Tokovinin}{2004}]{GLAO}
Tokovinin A.,  2004, PASP, 116, 941


\bibitem[\protect\citeauthoryear{Tokovinin et al.}{2003a}]{Rest}
Tokovinin  A., Kornilov V.,  Shatsky N.,  Voziakova O., 2003a,  
MNRAS, 2003, 343, 891


\bibitem[\protect\citeauthoryear{Tokovinin et al.}{2003b}]{Tok03}
Tokovinin A., Baumont S.,  Vasquez J., 2003b,  MNRAS,  340, 52


\bibitem[\protect\citeauthoryear{Tokovinin et al.}{2005}]{MASS-MK}
Tokovinin A., Vernin J., Ziad A.,  Chun M., 2005, PASP, 117, 395


\bibitem[\protect\citeauthoryear{Tokovinin \& Heathcote}{2006}]{Donut}
Tokovinin A., Heathcote S., 2006, PASP, 118, 1165 

\bibitem[\protect\citeauthoryear{Tokovinin \& Travouillon}{2006}]{TT06}
Tokovinin A.,  Travouillon T.,  2006, MNRAS, 365, 1235

\bibitem[\protect\citeauthoryear{Tokovinin \& Kornilov}{2007}]{TK07}
Tokovinin A.,  Kornilov V., 2007, MNRAS, accepted (TK07)

\bibitem[\protect\citeauthoryear{Wang et al.}{2006}]{TMT-DIMM}
Wang L., Schoeck M., Chanan G., 
Skidmore W., Bustos E., Seguel J., Blum R., 2006,
in  Stepp  L.M., ed., 
Ground-based   and  Airborne  Telescopes,  
Proc. SPIE, 6267, 62671S 


\bibitem[\protect\citeauthoryear{Ziad et al.}{2000}]{GSM}
Ziad A., Conan R., Tokovinin A., Martin F.,  Borgnino. J., 
2000,  Appl. Opt.,  39, 5415



\end{thebibliography}
\end{document}